\documentclass[a4paper,twocolumn]{esapub} 
\usepackage{natbib}
\usepackage{epsfig}

\title{SPI data analysis: the CESR/Toulouse approach}
\author{J\"urgen Kn\"odlseder}
\affil{Centre d'\'Etude Spatiale des Rayonnements, B.P. 4346, 31028 
Toulouse Cedex 4, France (knodlseder@cesr.fr)}

\begin{document}


\maketitle

\begin{abstract}

In order to allow for an efficient and flexible scientific analysis of data 
from the SPI imaging spectrometer aboard INTEGRAL, I developed a set of 
analysis executables that are publicly available through the internet. 
The software is fully compatible with the ISDC data format. 
It complements software that is actually available through ISDC, and will 
be included in the next ISDC software release. 
This paper describes the design of my software system and provides a brief 
introduction to the executables.
\end{abstract}

\section{Introduction}

SPI data analysis is a complex task. 
The data volume is important, the outstanding spectral resolution requires 
an accurate gain calibration, the important instrumental background demands 
a very precise modelling of its time variations, and the generally weak 
signals require sophisticated analysis methods. 
To perform SPI data analysis at the Centre d'\'Etude Spatiale des Rayonnements 
(CESR) in Toulouse, I therefore developed an efficient data analysis system 
that builds on the software kernel provided by the INTEGRAL Science Data 
Centre (ISDC). 
My software is available at the site 
{\tt http://www.cesr.fr/$\sim$jurgen/isdc/index.html} 
and will be distributed in future versions of the ISDC Offline Science Analysis 
(OSA) system. 
At CESR, the system has been installed on a UNIX machine under Solaris 9, 
yet ports to Linux systems have also been performed successfully (the 
installation procedure is identical to that of regular ISDC software).

The core of the system is a library of C++ classes and functions 
({\tt spi\_toolslib}) that provides all necessary functionalities for data 
analysis. 
Around this library, a number of analysis executables has been written. 
All software is documented by User Manuals. 
In particular, the {\tt spi\_toolslib} library is described in great detail, 
allowing the data analyst to easily build new analysis executables using the 
available functions.

The starting point of the analysis is the prepared data provided by ISDC. 
These data are organised by satellite orbital revolutions (one revolution lasting 
typically three days), and are split into so called science windows, which 
generally either comprise a spacecraft pointing or a spacecraft slew. 
In the current approach only the pointing data are exploited.

Data analysis consists of 5 steps:
observation group building,
gain calibration,
data preparation,
data combination,
scientific analysis (imaging, model fitting, spectral analysis)
In the following sections these 5 steps are described.

\section{Observation Group Building}

The entity of data that is combined for analysis is called an {\em observation 
group}. 
An observation group is a FITS file that contains pointers to all 
data that are associated to the observation.
For all executables it is sufficient to provide on input an 
observation group, the software then automatically extracts the 
relevant information.
Note, however, that the observation group contains the access paths 
to the individual data files, hence moving the data may invalidate the 
information stored in the observation group.
Thus it is recommended not to copy or move observation groups or 
associated data.

\begin{figure}[!t]
  \center
  \epsfxsize=3.5cm \epsfclipon
  \epsfbox{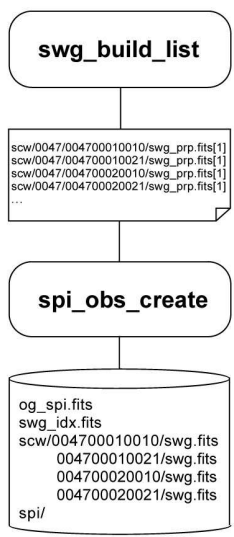}
  \caption{\label{fig:step1}
  Observation Group Building step.
  A list of science windows, specified by the Data Object Locations 
  (DOLs) of the science window groups listed in an ASCII file, are 
  combined into an observation group.
  This is the starting point for SPI data analysis.
  The resulting data structure is depicted in the data container at
  the bottom of the figure.
  }
\end{figure}

The first step of the analysis consists in building an observation group
(cf.~Fig.~\ref{fig:step1}).
For practical purposes, I generally build an observation group per revolution. 
This leads to manageable data sizes and reasonable execution times for data 
preparation. 
In addition, this allows for a time-dependent gain calibration that may 
compensate detector drifts.

To gather the available science windows of an observation group, I use 
a UNIX script called {\tt swg\_build\_list} that collects the {\em Data
Object Locations} (DOLs) of all {\em science window groups} into a single 
ASCII file.
A science window group is a FITS file that contains pointers to the 
data that are contained within an individual science window. 
The ASCII file is then used as input for the executable 
{\tt spi\_obs\_create} which builds the observation group.
One may edit the ASCII file to exclude individual science
windows from the analysis.
In the current system, a science window is the shortest junk of data
that may be analysed.
Analysis of shorter periods, e.g. as needed for gamma-ray burst 
studies, has not been implemented so far.

\section{Gain Calibration}

\begin{figure}[!ht]
  \epsfxsize=8cm \epsfclipon
  \epsfbox{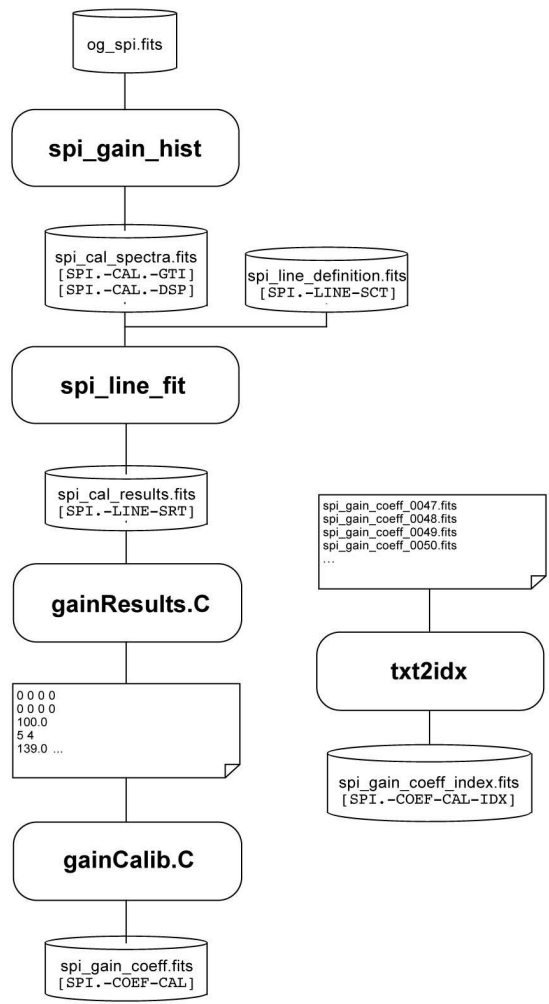}
  \caption{\label{fig:step2}
  Gain calibration step.
  The input observation group contains a list of science windows,
  typically those of one satellite revolution.
  The resulting gain coefficient file contains calibration 
  coefficients for this list of science windows.
  The coefficients for each revolution are combined into a
  calibration index allowing for time dependent gain
  correction.
  }
\end{figure}

The next step consists in gain calibration to convert the registered 
Pulse-Height Analyser (PHA) values for each event into physical meaningful 
energies (cf.~Fig.~\ref{fig:step2}).
For this purpose, the line centroid in PHA units are determined for a 
couple of well selected gamma-ray lines of known energies that arise in 
the instrumental background (see Lonjou et al., these proceedings). 
Spectra of PHA values are built for each revolution using the executable
{\tt spi\_gain\_hist}, and line fitting is performed using 
{\tt spi\_line\_fit}. 
In the actual approach, Gaussian shaped lines on top of a linear background 
are employed for fitting (this basically ignores the effect of detector 
degradation which leads to an extension on the left wing of the lines).

From the fitted energies, calibration relations are derived using the 
isdcroot scripts {\tt gainResults.C} and {\tt gainCalib.C}. 
The calibration relations for each revolution are then combined in an 
index table (using the ISDC executable {\tt txt2idx}) that is passed to 
the event histogramming software, which then handles the time dependence 
of the gain calibration.

Note that gain calibration files and indices are provided by 
ISDC and may also be downloaded on my internet site, hence normally
the gain calibration step is not required for your analysis.

\section{Data Preparation}

\begin{figure}[!t]
  \epsfxsize=8cm \epsfclipon
  \epsfbox{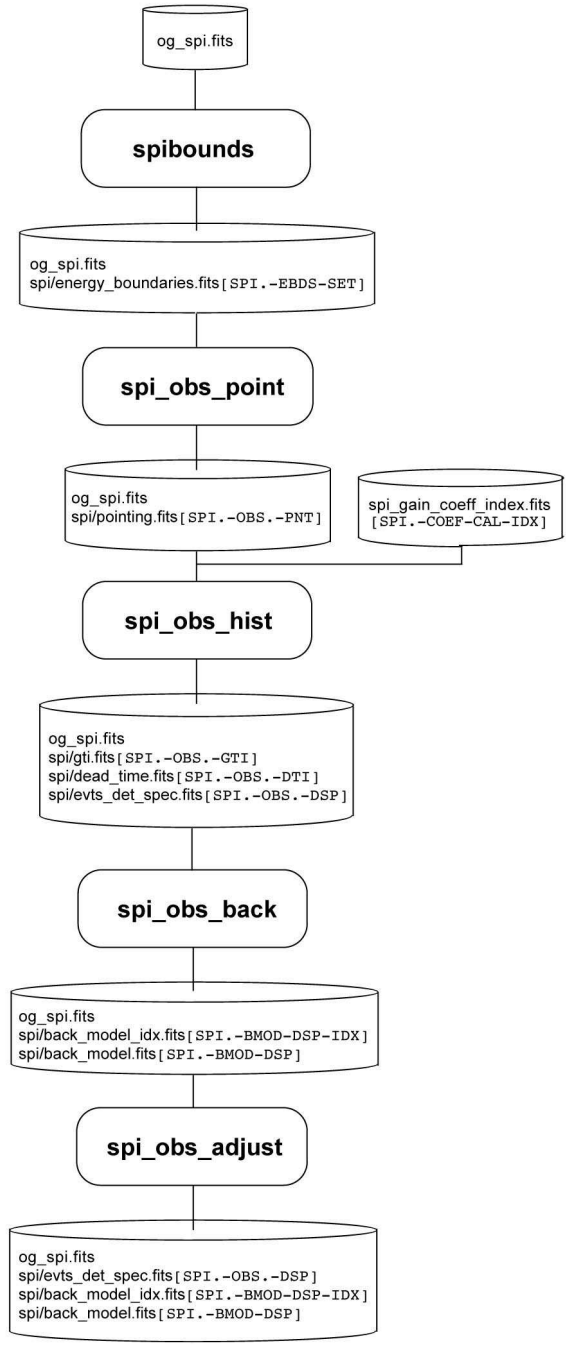}
  \caption{\label{fig:step3}
  Data preparation step.
  The input observation group contains a list of science windows,
  the output observation group contains calibrated detector spectra 
  for each pointing and pseudo-detector with the associated pointing, 
  exposure, lifetime and background model information.
  Degradation effects have been corrected for.
  The output observation group is now ready for scientific analysis.
  }
\end{figure}

The Data Preparation step consists in building the 3-dimensional SPI 
data-space for each observation group, which is spanned by spacecraft 
pointing, telescope detector, and photon energy  
(cf.~Fig.~\ref{fig:step3}). 
This data-space is stored in a FITS file with extension name 
{\tt SPI.-OBS.-DSP}.

First, the energy binning of the data-space is defined using the 
executable {\tt spibounds} which adds a FITS file with extension name 
{\tt SPI.-EBDS-SET} to the observation group that contains the lower and 
upper energy boundary for each of the energy bins.

As next step, the available spacecraft pointings are extracted for an 
observation group using the executable {\tt spi\_obs\_point}. 
This executable adds a FITS file with extension name {\tt SPI.-OBS.-PNT} 
to the observation group that contains the start and stop time for each 
pointing, as well as the SPI telescope pointing direction (the misalignment 
between SPI and the spacecraft is taken into account at this step).

Now the observation group is ready for data-space building. 
This is done using the executable {\tt spi\_obs\_hist} which adds three FITS 
files to the observation group: 
{\tt SPI.-OBS.-GTI} which contains the Good Time Intervals for each pointing 
and detector as well as the effective exposure time, 
{\tt SPI.-OBS.-DTI} which contains the Livetime for each pointing and detector, 
and {\tt SPI.-OBS.-DSP} which contains the histogrammed photon data (in units
of counts).
{\tt spi\_obs\_hist} automatically detects the data transmission modes 
of SPI (operational mode, spectral mode or emergency mode), and 
performs the required transformations to provide a homogeneous and 
clean set of prepared data. 

Before the data can be analysed, a model of the instrumental background is 
required. 
Background modelling is performed using the executable 
{\tt spi\_obs\_back} which adds a list of background model components 
to the 
observation group (extension name {\tt SPI.-BMOD-DSP}). 
Note that the background model may consist of several components which are 
stored in separate data structures (yet in the same FITS file). 
These data structures are then combined by an index table that is attached 
to the observation group (extension name {\tt SPI.-BMOD-DSP-IDX}).

Background modelling is of course the most delicate step in SPI data analysis, 
and for more information about the available background  models the data 
analyst is referred to the {\tt spi\_obs\_back} User Manual.

\begin{figure}[!t]
  \center
  \epsfxsize=5cm \epsfclipon
  \epsfbox{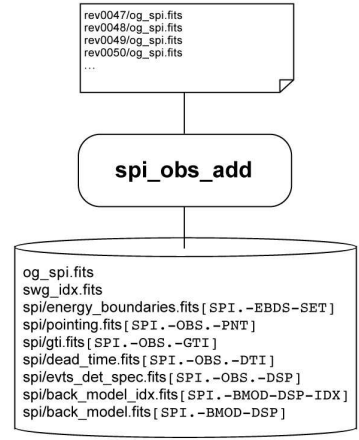}
  \caption{\label{fig:step4}
  Data combination step.
  A list of prepared observation groups are combined into a single
  observation group.
  Background modelling and degradation adjustment may be applied to the
  combined data.
  }
\end{figure}

Finally, differences in the detector degradation between the histogrammed 
data and the background model components may be compensated using the 
executable {\tt spi\_obs\_adjust}. 
This step is optional, but recommended for gamma-ray line analysis.

\section{Data Combination}

If the data preparation has been performed revolution-wise, the observation 
groups may now be combined using the executable {\tt spi\_obs\_add}
(cf.~Fig.~\ref{fig:step4}). 
Note that in principle several revolutions may be prepared in a single step, 
yet the execution time of the histogramming step ({\tt spi\_obs\_hist}) 
rises non-linearly with data-space dimensions (due to the particular 
architecture of the FITS access routines).

{\tt spi\_obs\_add} also allows to select a sub-range of energy bins
from the input data, leading eventually to smaller and more manageable 
data-space dimensions.
Background modelling ({\tt spi\_obs\_back}) and degradation adjustment 
({\tt spi\_obs\_adjust}) may also be performed after data combination 
using the output observation group of {\tt spi\_obs\_add} as input.
This allows testing different background models and/or degradation 
corrections for the same data.

\section{Scientific Analysis}

\begin{figure}[!t]
  \epsfxsize=8cm \epsfclipon
  \epsfbox{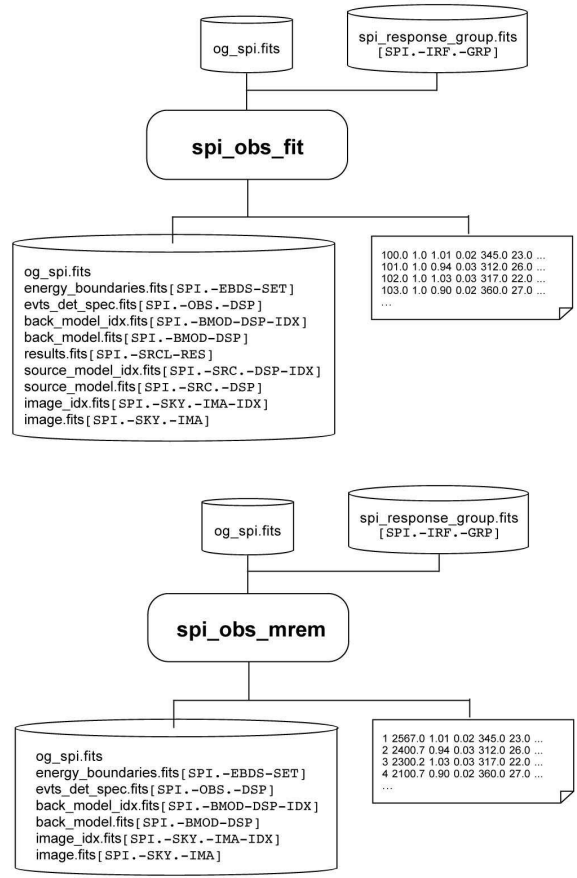}
  \caption{\label{fig:step5}
  Scientific analysis step.
  The instrument response is used in this step.
  On output, an observation group is created that contains all data
  that have been used for the scientific analysis.
  An ASCII file contains the analysis results for quick display.
  }
\end{figure}

Now, scientific analysis executables provided by ISDC, such as {\tt spiros} 
for point-source localisation and spectral analysis, may be applied to the 
prepared data.
In my system, two further executables are available
(cf.~Fig.~\ref{fig:step5}). 
{\tt spi\_obs\_fit} performs fitting of sky-intensity distributions to the 
data, allowing for morphology studies and spectral analysis. 
Various treatments of the instrumental background are provided, allowing 
for reduction of the systematic uncertainties in the data analysis.

{\tt spi\_obs\_mrem} performs Expectation Maximisation image deconvolutions 
to extract sky-intensity distributions from the data. 
The implemented algorithm is also known as accelerated Richardson-Lucy 
algorithm, and has proven successful for the analysis of gamma-ray data 
for other missions. 
In the future, it is foreseen to implement also a Multiresolution analysis 
in this software, allowing to reduce statistical artefacts in the 
reconstructed intensity distributions for diffuse emission.

Both executables produce on output an observation group that contains 
the data that have been used for analysis.
This means that if for example only a single energy bin has been selected for 
imaging analysis, the resulting observation group contains only the 
data and background model for this single energy bin.
Or if the background model has been modified by {\tt spi\_obs\_fit} 
during the fitting procedure, the resulting background model with the 
scaling factors applied will be stored in the output observation group.
In a following step, this output observation group may then be used 
for example for imaging.

{\tt spi\_obs\_fit} provides also an image on output which contains 
the best fitting combination of the model components that have been 
fitted to the data.
In this way, a model intensity distribution can be derived from the 
data.

\section{Conclusions}

The software described in this article has been designed and coded to 
allow for an efficient and flexible scientific analysis of SPI
data.
I hope that the present article is particularly useful for INTEGRAL 
guest observers, and that the information provided will increase the 
reliability of the data analysis performed by non experts.
Providing simple and robust software to the user community is 
certainly a prerequisite for the success of the INTEGRAL mission, 
hopefully allowing to achieve many new and interesting science results.

\end{document}